\documentclass[sigconf]{acmart}
\usepackage{balance}
\setcopyright{none}
\usepackage{balance}
\usepackage[colorinlistoftodos,prependcaption,textsize=tiny]{todonotes}
\usepackage{caption}
\usepackage{tabularx, environ}
\usepackage{titlesec}
\usepackage{float}
\usepackage{ifsym}

\copyrightyear{2021}
\acmYear{2021}
\setcopyright{acmcopyright}\acmConference[SIGIR '21]{Proceedings of the 44th International ACM SIGIR Conference on Research and Development in Information Retrieval}{July 11-15, 2021}{Virtual Event}
\acmBooktitle{Proceedings of the 44th International ACM SIGIR Conference on Research and Development in Information Retrieval (SIGIR '21), July 11-15, 2021, Virtual Event}
\acmPrice{15.00}

\begin{document}
\fancyhead{}

\title{APRF-Net: Attentive Pseudo-Relevance Feedback Network for Query Categorization}

\author{Ali Ahmadvand}
\affiliation{Emory University\country{USA}}
\email{ali.ahmadvand@emory.edu}

\author{Sayyed M. Zahiri}
\affiliation{\institution{The Home Depot\country{USA}}}
\email{mzahiri@gatech.edu}

\author{Simon Hughes}
\affiliation{\institution{The Home Depot\country{USA}}}
\email{simon_hughes@homedepot.com	}

\author{Khalifa Al Jadda}
\affiliation{\institution{The Home Depot\country{USA}}}
\email{khalifeh_al_jadda@homedepot.com	}

\author{Surya Kallumadi}
\affiliation{\institution{The Home Depot\country{USA}}}
\email{surya@ksu.edu}

\author{Eugene Agichtein}
\affiliation{Emory University\country{USA}}
\email{eugene.agichtein@emory.edu}

\begin{abstract}
  Query categorization is an essential part of query intent understanding in e-commerce search. A common query categorization task is to select the relevant fine-grained product categories in a product taxonomy. For frequent queries, rich customer behavior (e.g., click-through data) can be used to infer the relevant product categories. However, for more rare queries, which cover a large volume of search traffic, relying solely on customer behavior may not suffice due to the lack of this signal. To improve categorization of rare queries, we adapt the Pseudo-Relevance Feedback (PRF) approach to utilize the latent knowledge embedded in semantically or lexically similar product documents to enrich the representation of the more rare queries. To this end, we propose a novel deep neural model named \textbf{A}ttentive \textbf{P}seudo \textbf{R}elevance \textbf{F}eedback \textbf{Net}work (APRF-Net) to enhance the representation of rare queries for query categorization. To demonstrate the effectiveness of our approach, we collect search queries from a large commercial search engine, and compare APRF-Net to state-of-the-art deep learning models for text classification. Our results show that the APRF-Net significantly improves query categorization by 5.9\% on $F1@1$ score over the baselines, which increases to 8.2\% improvement for the rare (tail) queries. The findings of this paper can be leveraged for further improvements in search query representation and understanding. 
  
\end{abstract}


\maketitle

\section{Introduction and Related Work}

In the realm of Web and e-commerce search, query categorization is an important step in query intent understanding, which in e-commerce can be viewed as mapping search queries to relevant fine-grained product categories \cite{zhao2019dynamic}. It has been shown that query categorization could play a pivotal role in increasing user satisfaction by returning more relevant products in e-commerce search \cite{zhang2019generic}. Query categorization is challenging due to multiple reasons: (i) queries are generally short and might suffer from lack of concrete evidence of customers' intents \cite{montazeralghaem2020reinforcement}, (ii) customers have different ways to express the same intent. For example, ``blocktile white-colored - 12 in. x 12 in.'' and ``block tile white 12x12 inches'' (iii) query categorization is an extreme multi-label text classification problem (XMTC), which raises challenges such as data sparsity and scalability \cite{liu2017deep, wang2018joint, bojanowski2017enriching}, and (iv) the most challenging case is handling rare queries (a.k.a. tail queries). Tail queries suffer from the lack of customer behavior signals (e.g., click-through data), which could lead to poor representation of rare queries. Rare queries are generally caused by typos, synonyms, morphological variants, and, more importantly, when customers express their intent in a unique fashion \cite{yin2016ranking}. Different methodologies have been proposed to deal with tail queries in retrieval systems, such as corpus-based and knowledge-based query expansion methods \cite{lavrenko2017relevance} and \cite{montazeralghaem2020reinforcement}. Although standard query expansion models are effective, they might not be sufficient since even perfect expansion terms for a tail query might produce another tail query \cite{song2014transfer}. Also, adding new terms in query expansion may result in a higher recall but, sometimes a lower precision, and (ii) the suggested terms are not ordered, which could be problematic for text embedding models trained on natural language text. Unlike standard query expansion techniques, our model is trained to attend across document fields based on their semantic similarities with a query, which, as will show, improves the performance on the query categorization task.

In this paper, inspired by \cite{li2018nprf}, we propose a deep learning-based PRF model, named Attentive Pseudo-Relevance Feedback Network (APRF-Net). Unlike \cite{li2018nprf} that was developed to boost ranking performance, APRF-Net is an extreme multi-label classifier that is specifically developed for e-commerce query categorization. To mitigate tail queries' sparsity, APRF-Net utilizes PRF to enrich the tail queries' representations with the information in the retrieved product documents. The product documents in e-commerce websites are well-structured and contain fields like title, short description, color, etc. In our setting, fields form product documents, and top-ranked documents create a corpus. To capture this existing hierarchical structure, APRF-Net jointly generates query and product field-level embeddings and utilizes hierarchical attention to prioritize the top-k retrieved product documents at three levels of abstractions (field, document, and corpus). In summary, our main contribution is proposing a new model to address the customer signal gap between frequent and rare queries by leveraging the informative signals from PRF for query categorization in e-commerce.




\begin{figure*}
\centering
\includegraphics[width = 500pt]{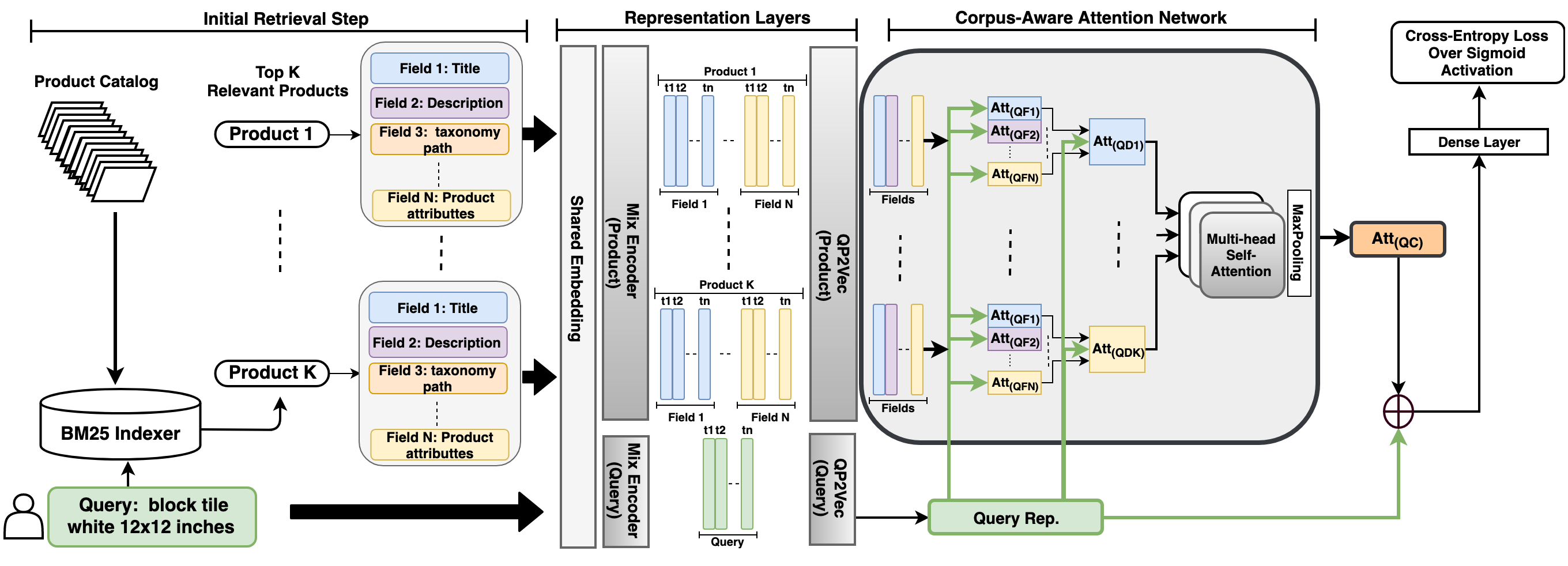}
\caption{Architecture of the proposed attentive pseudo-relevance feedback network (APRF-Net).}
\label{fig:Model}
\end{figure*}

\begin{figure}
\centering
\includegraphics[width = 250pt]{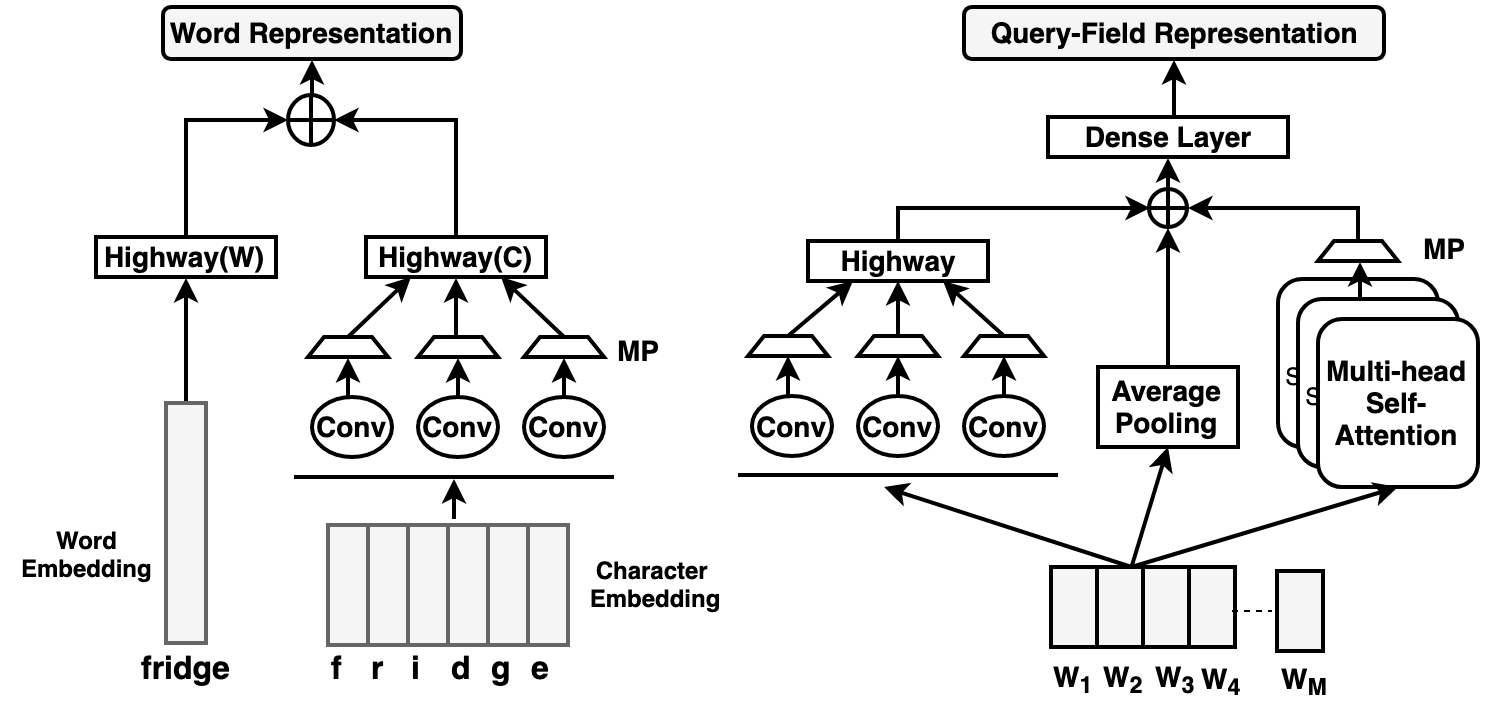}
\caption{(a) Mix Encoder layer (b) QP2V layer.}
\label{fig:QP2Vec}
\end{figure}

\section{Model Overview}

APRF-Net aims to map a query to one or multiple product categories by incorporating its top-ranked retrieved products' content. As illustrated in Figure \ref{fig:Model}, APRF-Net is comprised of three components:

\begin{enumerate}

   \item \textbf{Initial Retrieval Step:} This component is devised to return top-k relevant product documents for an issued query. 
    \item \textbf{Representation Layers:} These layers jointly learn sophisticated query and product field context vectors (Figure. \ref{fig:QP2Vec}).
    \item \textbf{Corpus-Aware Attention Network:} A hierarchical attention network that models the hierarchical structure of the top retrieved product documents (field, document, and corpus). 
    
\end{enumerate}

\subsection{Initial Retrieval Step} 

APRF-Net uses BM25 model to retrieve the top-k relevant product documents (corpus) \textbf{$C_k=\{d_1,...,d_k\}$} for an initial query \textit{q}. Each product document is broken down into its corresponding fields (title, color,...). Finally, The query and its corresponding documents' fields are fed into the representation layers (subsection \ref{sssec:rep layer}).

\subsection{Representation Layers} \label{sssec:rep layer}
The objective of these layers is to generate context vectors for a query and its retrieved product documents' fields. Representation Layers consist of the following layers: (i) Embedding  (ii) Mix Encoder (iii) Query-product-to-vector (QP2Vec). All the layers above are shared between queries and products. A shared embedding could bridge the vocabulary gap between queries and documents. Also, shared encoders enable the network to transfer the captured knowledge across both product documents and queries.

\subsubsection*{\bf Mix Encoder Layer.}

Tail queries generally contain many infrequent terms due to different reasons, such as typos. To make the model less susceptible to out-of-vocabulary issues and spelling errors, we enrich word-level representations with character-aware embeddings introduced by \cite{zhang2019generic}.

\subsubsection*{\bf Query-Product-to-Vector Layer (QP2Vec):} QP2Vec composes enhanced word representations from mix encoder to generate a query or products' field-level representations (Figure \ref{fig:QP2Vec}.b). The network takes word embeddings and passes them into three parallel neural layers. (i) A three-layer CNN model followed by a max-pooling layer, (ii) an average pooling layer, and (iii) a multi-head self-attention network. Multi-head self-attention contains several linear projections of a single scaled dot-product function that are parallelly implemented. Eq. \ref{head} shows a single head in self-attention.
\begin{equation}
        head_t = softmax \Big( \frac{ E_i K^T}{\sqrt{d_k}} \Big) \mathrel{V}, t=0,...,h
        \label{head}
\end{equation}
Where $K$ is the key matrix, $V$ is the value matrix, and $d_k$ is the dimension of the keys, $E_i$ is word embedding for $i-th$ token generated by the mix encoder. Finally, heads' outputs are concatenated and multiplied to a linear transformation, and passed to a max-pooling layer to generate the final representation (Eq. \ref{eq:self}).
\begin{equation}
    SelfAttention(E) = MaxPooling([head_1; ...;head_h] . W)
    \label{eq:self}
\end{equation}

Where $W$ is the weights of a linear function. All the three outputs from previous steps are concatenated for final representations of query or product's field (Eq. \ref{eq:QP}).

\begin{equation}
        QP = [Avg(E); SelfAttention(E);CNN(E)], QP \in R^F
        \label{eq:QP}
\end{equation}

Where $F$ is the dimension of the query or fields. The $QP$ vectors are sent to the corpus-aware attention network (section. \ref{CAAN}).

\subsection{\bf{Corpus-Aware Attention Network}}
\label{CAAN}
It is a hierarchical attention network which enriches query representations by incorporating informative contents from top retrieved documents. Corpus-aware attention network contains three levels of abstractions: (i) \textbf{query-field attention layer:} we apply a separate query-field attention operation, $Att_{QF_i}$, across all document fields $i\in\{1,..,N\}$ and top-K retrieved documents. As a result of this operation, $(K \times N)$ vectors of size $(1 \times F)$ are produced, where $N$ is the number of document fields and $K$ is the number of top retrieved documents. To form document attention representation $Att_{D_k} \in R^{N \times F}$, we stack document's query-field attentions: $Att_{QF[k \times N : (k+1) \times N]}$ where $k\in\{0,..,K-1\}$, (ii) \textbf{query-document attention layer:} $Att_{D_k}$ matrices are fed into another attention layer to generate query-document attention matrix $Att_{{QD}}$ (Eq. \ref{eq:QD}). To implement the attention layers for both (i) and (ii), we used Loung-style attention \cite{luong2015effective}.

\begin{equation}
    Att_{QD} = [ Q \odot Att_{D_0};...;Q \odot Att_{D_K} ], Att_{{QD}} \in R^{K \times D} 
    \label{eq:QD}
\end{equation}
 Where $D$ is the dimension of query-document attention.
\noindent(iii) \textbf{query-corpus attention layer:} the query-document attentions are passed into a self-attention with $K$ heads followed by a max-pooling to generate query-corpus attention $Att_{QC}$ (Eq. \ref{eq:QC}). 

\begin{equation}
   Att_{QC} = Maxpooling(selfattention(Att_{QD})), Att_{QC} \in R^{1 \times C} 
   \label{eq:QC}
\end{equation}

Where $\odot$ and $;$ indicate the attention layer and concatenation, respectively. $Att_{QC}$ is the final PRF signal from top-K returned documents to expand the query representations. To do so, we concatenate $Att_{QC}$ to the original query representation. In other words, we expand the query model using the document model in the latent space. Finally, the output is fed into a dense layer to fuse query and document representations before the final category prediction.

\section{DATASET OVERVIEW}
\label{sec:Dataset}

This section describes our dataset creation process from e-commerce search logs. We use six months of search data to create train, validation, and test sets.  Similar to \cite{zhao2019dynamic}, we utilize click-through data to approximate ground truth labels (product categories) by considering the following steps: for a query, (i) leverage query-product clicks to find candidate categories, (ii) aggregate the number of clicks across these categories, and (iii) assign only categories with more than 5\% of total clicks. The most fine-grained product categories located in the leaf nodes of taxonomy paths are considered as final classes (e.g., ``Bath>Bathroom Faucets>Bathroom Sink Faucets>Single Hole Bathroom Faucets'' -> ''Single Hole Bathroom Faucets''). For each query, we also collect the top-ranked product documents. The table below summarizes fields included in a product document. We randomly sample 60\% of the data to create the training, 10\% for validation, and the remaining 30\% for test sets. We collect 60M search queries (2.5M unique queries) that contain customer signal information, including 4,665 unique product categories.

\begin{center}
\small
    \begin{tabular}{l|l}
    \toprule
    \bf  Fields & \bf Short Description \\
    \bottomrule
        Title & Product Title \\
        Description & Product descriptions \\
        Taxonomy path & Product taxonomy path from root to leaf\\
        Color/material &  Color and material information \\
        Numerical & Numeric values (height, width, etc.) \\
        Brand & Brand information (brand name) \\

    \bottomrule
    \end{tabular}
    \label{tab:document}
\end{center}

We segment the test set into two types: (i) \textbf{Traffic Type (TT):}  we utilize the query's frequency information from one year (the refined queries are used to compute the query frequency) with a certain threshold to map each test query to a traffic bucket (head, torso, and tail). The distributions of head, torso, and tail queries in the Traffic Type test set are 8\%, 62\%, and 30\%, respectively. (ii) \textbf{Query Type (QT):} queries could contain one or several attributes. We randomly sample 10k queries from the initial test set to assign them to these five attributes: brand name, product name, ID (model number, universal product code, and serial number), numerical (dimension, units, etc.), and color/material. Note that the aforementioned query types are not mutually exclusive, i.e., a query could be mapped to multiple attributes. We leverage some rule-based algorithms to create the initial query type; later, the ones with low confidence scores are relabeled by human annotators.

\begin{table*}
\small
\centering
\begin{tabular}{@{}l|lll|lll|lll|l@{}}
\toprule  
\bf Method& \bf P@1&\bf R@1&\bf F1@1&\bf  P@2&\bf R@2&\bf F1@2&\bf P@3&\bf R@3&\bf F1@3&\bf MAP@3 \\
\toprule

CNN+SCE&0.540&0.486&0.511&0.327&0.574&0.416&0.245&0.633&0.353&0.371\\
BiLSTM+SCE&0.546&0.419&0.448&0.317&0.554&0.403&0.241&0.619&0.347&0.367\\
FastText &0.571&0.450&0.503&0.366&0.576&0.448&0.271&0.639&0.380&0.403\\
XML-CNN &0.564&0.507&0.539&0.339&0.593&0.431&0.252&0.650&0.363&0.385\\
LEAM &0.578&0.518&\underline{0.546}&0.344&0.601&\underline{0.438}&0.253&0.652&\underline{0.365}&\underline{0.391}\\
\hline

\hline

APRF-Net, top-K=1 &$0.599^*$&$0.540^*$&$0.567^*$&$0.353^*$&$0.622^*$&$0.450^*$&$0.260^*$&$0.676^*$&$0.376^*$&$0.404^*$\\
APRF-Net, top-K=3&\bf$0.610^*$&$0.548^*$&\bf$0.578^*(+5.9\%)$&\bf$0.361^*$&\bf$0.633^*$&\bf$0.460^*(+5.0\%)$&\bf$0.265^*$&\bf$0.6879^*$&\bf$0.383^*(+4.9\%)$&\bf$0.412^*(+5.4\%)$\\

\bottomrule
\end{tabular}
\caption{Performances of query categorization on TT test set. ``*'' indicates statistically significant improvements $\textit{p} < 0.05. (X\%)$: shows relative improvement between bolded and underlined scores. P and R stand for precision and recall. }
\label{tab:overall}
\end{table*}

\section{EXPERIMENTAL SETUP}
\label{sec:es}
This section describes the parameter settings, metrics, baselines, results, and experimental procedures used to evaluate our model.
. 
\subsubsection*{\bf Parameter Settings.} An Adam optimizer was used with a learning rate of $\eta=0.001$, a dropout rate of 0.5, a mini-batch of size 64, and embedding sizes of 300 and 16 for word and characters, respectively. For convolutional layers, we employed 128 filters with kernel sizes of 1, 2, and 3. We utilized Word2Vec to initialize the word embeddings, and character embeddings were initialized randomly. We employed Sigmoid Cross-Entropy (SCE) for the loss function as it shows superior results compared to the rank loss for XMTC problems \cite{nam2014large}. All models were implemented by Tensorflow 2.1 with a single NVIDIA P100 GPU.

\subsubsection*{\bf Evaluation Metrics. }

Following the conventions of the literature for XMTC problems and specifically in product categorization \cite{zhao2019dynamic}, we reported Precision@K ($P@K$), Recall@K ($R@K$), $F1@K$, and Mean Average Precision@K ($MAP@K$) on the top-K results. 


\subsubsection*{\bf Methods Compared. }
\label{sec:comp}
We employed the following deep learning models as baselines to evaluate the effectiveness of APRF-Net.   
\begin{itemize}
    \item \textbf{CNN+SCE:} 3-layers convolutional neural network \cite{kim:2014}.
    \item \textbf{BiLSTM+SCE:} Bidirectional recurrent neural network.
    \item {\bf FastText:} Bag of tricks for efficient text classification\cite{bojanowski2017enriching}. 
    \item {\bf XML-CNN:} Deep learning for XMTC \cite{liu2017deep}.
    \item {\bf LEAM:} Joint label embedding attentive model \cite{wang2018joint}.
    \item {\bf APRF-Net:} The proposed model.
    
\end{itemize}
\section{Empirical Results and Discussion}
Table. \ref{tab:overall} summarizes the performances of models in section. \ref{sec:comp}. As shown in Table \ref{tab:overall}, APRF-Net achieved 5.9\% relative improvement on $F1@1$ compared to LEAM, the best performing baseline $@1$. As $K$ increases, $P@K$ scores drop since each query on average has 1.26 relevant categories (head : 3.3, torso : 1.2, and tail : 1.0). 

\subsubsection*{\bf Results on Traffic and Query Types.} Table \ref{tab:buckets} shows the performance of APRF-Net compared to LEAM on different buckets. APRF-Net advantages are promising, stable, and stronger consistently across all metrics and buckets. We can conclude that APRF-Net has a more significant impact on rare queries across all metrics from our experimental results. For instance, in terms of $F@1$ score, it achieves (3.8\%, 5.2\%, and 8.2\%) relative improvements on (head, torso, and tail), compared to LEAM, respectively. This result proves our earlier claim that our proposed model can make the tail queries less sparse by transferring knowledge from semantically similar documents across queries, specifically, from frequent to rare queries. 

\begin{table}
\small
\centering
\begin{tabular}{@{}l|l|l|l|l|l|l@{}}
 \toprule 
\bf Test& \bf Bucket&\bf Method& \bf F1@1&\bf F1@2&\bf F1@3&\bf MAP@3 \\
\toprule
  \bf TT&head&LEAM & 0.444& 0.428& 0.412& 0.567\\
  &torso&LEAM & 0.541& 0.426& 0.351& 0.384\\
  &tail&LEAM &0.539 & 0.418& 0.339& 0.362\\

\hline
 \bf TT&head&APRF-Net,3 &0.461&0.436&0.414&0.576\\
            &torso&APRF-Net,3 &0.569&0.445&0.367&0.402\\
            &tail&APRF-Net,3 &0.583&0.449&0.365&0.391\\
  \bottomrule \bottomrule
  \bf Test& \bf Bucket&\bf Method& \bf F1@1&\bf F1@2&\bf F1@3&\bf MAP@3 \\
  \bottomrule
  \bf QT&brand&LEAM &0.605 & 0.475& 0.391& 0.437\\
             &product&LEAM & 0.604& 0.477& 0.395&0.442 \\
             &num&LEAM & 0.663& 0.503& 0.409& 0.456\\
             &ID&LEAM &0.174 &0.140 &0.114 &0.122 \\
             &c/m&LEAM & 0.730& 0.548&0.439 & 0.498\\
\hline
 \bf QT&brand&APRF-Net,3 &0.641&0.500&0.410&0.460\\
             &product&APRF-Net,3 &0.626&0.495&0.409&0.457\\
             &num&APRF-Net,3 &0.701&0.535&0.427&0.482 \\
             &ID&APRF-Net,3 &0.241&0.191&0.162&0.441\\
             &c/m&APRF-Net,3&0.744&0.546&0.441&0.503\\
\bottomrule
\end{tabular}
\caption{Performances across buckets. c/m:color/material}
\label{tab:buckets}
 \vspace{-0.8cm}
\end{table}

Results on the QT test set are also promising. In particular, the improvement is significant for queries with ID attributes (with 38\% relative improvement on $F1@1$ over LEAM). This low performance is caused by the lack of natural language content in ID queries, which can be alleviated by including PRF signals.

\subsubsection*{\bf Results on Minority Classes.}
Due to the intrinsic imbalanced nature of e-commerce query categorization, our training data exhibits a power-law distribution. We used the method introduced by \cite{jiang2012scaling} to select the minority classes, then evaluate APRF-Net on the picked minority categories. As a result, 3,130 out of 4,665 categories are considered as minority classes in our dataset. The LEAM achieved (0.293\%, 0.257\%, 0.226\%, and 0.222\%), and APRF-Net reached (0.350\%, 0.295\%, 0.257\%, and 0.256\%) on ($F1@1$, $F1@2$,$F1@3$, and $MAP@3$), respectively. The results show a 19.5\% $F1@1$ relative improvement over LEAM.

\subsection{\bf Ablation Analysis }
APRF-Net is a complex model that consists of several components: shared embedding, mix encoder, QP2Vec, and corpus-aware attention network. We performed a comprehensive ablation study to evaluate each component's impact on the overall performance (Table. \ref{tab:ablation}). The results illustrate that by utilizing PRF signals, APRF-Net achieved relative boosts by (8.0\%, 5.4\%, 3.9\%, and 8.9\%) compared to QP2Vec on ($F1@1, F1@2, F1@3,$ and $MAP@3$), respectively.

\begin{figure}
\centering
\includegraphics[width = 250pt]{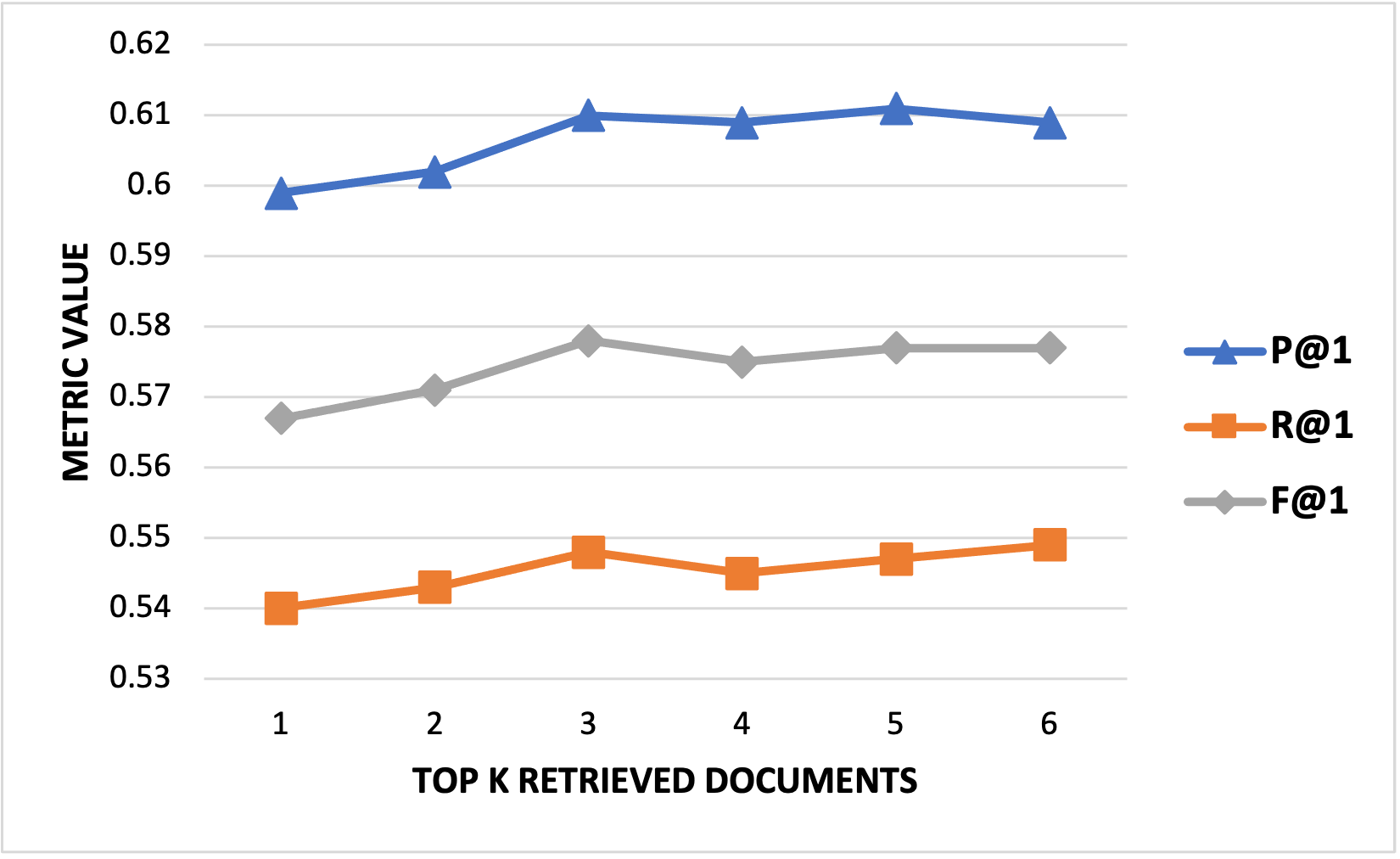}
\caption{Impact of top-K documents on APRF-Net.}
\label{fig:topk}
\end{figure}

To analyze the impact of the number of retrieved documents on APRF-Net model, we ran experiments with different top-K (document) values. As shown in Figure. \ref{fig:topk}, there is a sharp increase from 1 to 3, and it plateaus out afterward. This is due to two reasons: (i) after 3, the relevancy of documents drops, and (ii) the model has already captured the informative contents. Thus, we chose 3 the optimal value for our model.

\subsubsection*{\bf Comparing Corpus-Aware Attention Network with RM3\footnote{We used Lucene with default parameters available at \url{https://github.com/castorini/pyserini}}.} To further investigate the robustness of the corpus-aware attention network for expanding queries in a latent space, we compared it with standard PRF models (e.g., RM3 \cite{lavrenko2017relevance}) that expand queries at the term-level. To do so, first, we expanded the queries using the top 10 relevant documents to [query <\*expand> new terms] utilizing RM3, then, a multi-label classifier such as QP2Vec was used for final classification. QP2Vec+RM3 obtained (0.548, 0.442, 0.371, and 0.394) on ($F1@1, F1@2, F1@3,$ and $MAP@3$), which showed 4.4\% relative improvement on $F1@1$ compared to QP2Vec. Although RM3 provided improvements on all metrics compared to QP2Vec, we experienced a more significant improvement when we expanded the queries in a latent space using APRF-Net (Table. \ref{tab:ablation}).

\section{CONCLUSIONS}
We introduced APRF-Net, which adapts the idea of pseudo-relevance feedback (PRF) for the query categorization task in e-commerce search, especially to improve performance for rare queries. We proposed APRF-Net, a novel corpus-aware attention neural model to incorporate the PRF information, which represents a query using three abstraction levels (e.g., fields, documents, and corpus) on top-K retrieved products. The goal was transferring customer signal information from head queries to tail queries by leveraging semantic knowledge shared in the overlapping or similar retrieved product documents. Our results demonstrated the APRF-Net significantly improved query categorization by 5.9\% relative improvement on $F1@1$ score over the baseline, particularly 8.2\% improvement for tail queries. We plan to release the implementation of our model to serve as a benchmark for the research community. 

\begin{table}
\small
\centering
\begin{tabular}{@{}l|l|l|l|l@{}}
 \toprule 
\bf Method& \bf F1@1&\bf F1@2&\bf F1@3&\bf MAP@3 \\
\toprule
QP2Vec & 0.525&0.427& 0.362&0.371\\
QP2Vec+MixEncoder&0.536\tiny(+2.1\%)& 0.431\tiny(+0.1\%)& 0.366\tiny(+0.1\%)&0.383\tiny(+2.2\%)\\
APRF-Net,1 w/o SE &{0.553\tiny{(+5.3\%)}}&0.441\tiny(+3.2\%)& 0.369 \tiny(+1.9\%)&0.396\tiny(+6.7\%)\\
APRF-Net,1 &{0.567\tiny{(+8.0\%)}}&0.450\tiny(+5.4\%)& 0.376 \tiny(+3.9\%)&0.404\tiny(+8.9\%)\\

APRF-Net,3 &\bf{0.578\tiny\textbf{(+10.1\%)}}&\bf0.460\bf\tiny(+7.7\%)& \bf0.383 \tiny(+5.8\%)&\bf0.412\tiny(+11.1\%)\\

\bottomrule
\end{tabular}
\caption{Impact of APRF-Net components: QP2Vec, added by MixEncoder, corpus-aware attention network, and shared embedding. SE stands for the shared embedding.}
\label{tab:ablation}
 \vspace{-0.4cm}
\end{table}

\subsubsection*{\bf Acknowledgements.}
We would like to thank Faizan Javed, the Senior Manager of Search, for his valuable insights and feedback for this paper.


\balance

\bibliographystyle{abbrv}
\bibliography{References}

\end{document}